\documentclass[twocolumn,showpacs,preprintnumbers,amsmath,amssymb,prd]{revtex4}

\usepackage{graphicx}
\usepackage{dcolumn}
\usepackage{bm}
\usepackage{color}

\newcommand{\eref}[1]{Eqs.~(\ref{#1})}
\newcommand{\sref}[1]{section~\ref{#1}}
\newcommand{\cref}[1]{chapter~\ref{#1}}
\newcommand{\fref}[1]{figure~\ref{#1}}
\newcommand{\tref}[1]{table~\ref{#1}}

\newcommand{\Cref}[1]{Chapter~\ref{#1}}

\newcommand{\beq}[1] {\begin{equation} \label{#1}}

\newcommand{\eeq} {\end{equation}}
\newcommand{\noi} {\noindent}

\begin{document}


\title{Sensitivity  of  the   spherical  gravitational  wave  detector
MiniGRAIL operating at 5 K.}

\author{L. Gottardi}

 \altaffiliation[Current  address:  ]{  SRON, National  Institute  for
Space  Research,  High  Energy  Astrophysics  Division,  Utrecht,  the
Netherlands}
\email{l.gottardi@sron.nl} 
\author{A.  de  Waard} \author{A.  Usenko}
\author{G.  Frossati} 
\affiliation{LION,   Institute  of   Physics,
Kamerlingh   Onnes  Laboratorium,   Leiden  University,   Leiden,  The
Netherlands} 
\author{M.  Podt} 
\altaffiliation[Current   address:
]{Thales     Nederland     B.V.,     Hengelo,    the     Netherlands.}
\author{J. Flokstra} 
\affiliation{Low Temperature Division, Faculty of
Science and Technology,  Twente University, Enschede, The Netherlands}
\author{M.    Bassan}   \author{V.   Fafone}    \author{Y.   Minenkov}
\author{A.  Rocchi} 
\affiliation{Dip.  Fisica, Universitá  Tor Vergata
and INFN Roma2, Roma, Italy}

\date{\today}

\begin{abstract}
We present the performances and the strain sensitivity of the first
spherical gravitational wave detector equipped with a capacitive
transducer and read out by a low noise two-stage SQUID amplifier and
operated at a temperature of 5 K.  We characterized the detector
performance in terms of thermal and electrical noise in the system
output sygnal. We measured a peak strain sensitivity of $1.5\cdot
10^{-20} Hz^{-1/2}$ at 2942.9 Hz.  A strain sensitivity of better than
$5\cdot 10{-20}Hz{-1/2}$ has been obtained over a bandwidth of 30
Hz. We expect an improvement of more than one order of magnitude when
the detector will operate at 50 mK.  Our results represent the first
step towards the development of an ultracryogenic omnidirectional
detector sensitive to gravitational radiation in the 3kHz range. 

\end{abstract}

\maketitle

\section{INTRODUCTION}

The direct observation of gravitational waves (GWs) is one of the most
challenging tasks for experimental physics. After the first detection
will be claimed a new branch of astronomical observation will begin
and gravitational wave observatory will become more and more common
facilities. A spherical detector is a perfect instrument for an
astronomical observatory due to its feature of omnidirectionality and
polarization sensitivity \cite{Coccia95,Lobo95,Merkowitz98,
ZhouMich95, Stevenson98}.  The first ultracryogenic spherical
gravitational wave detectors \cite{MINIGRAIL,SCHENBERG} are currently
completing their engineering phase and will soon be operational with
an expected sensitivity better than $10^{-21} Hz^{-1/2}$ at $3 kHz$.
We report the results of the first sensitive measurement of the
spherical gravitational wave detector Minigrail.  The detector
read-out is based on capacitive resonant transducers coupled to a
superconducting quantum interference device (SQUID) linear amplifiers
by means of superconducting transformers. The two-stage SQUID
described in this work is one of the most sensitive amplifier ever
used on a gravitational wave antenna. It consists on a sensor dc SQUID
amplified by a Double Relaxation Oscillation Squid (DROS). The
coupling of the SQUID system to the high Q electrical resonators is
similar to the one developed for the AURIGA detector \cite{Vinante02}
We obtained a coupled additive energy resolution of $700 \hbar$ at $5
K$ in agreement with the expected values calculated from the SQUID
parameters using the standard model \cite{TESCLA77}.  The spherical
antenna described in this paper is the first example of a multimodal
resonant detector where the five quadrupolar modes of the sphere are
read-out by three resonant transducers.  In this paper we discuss the
noise contribution and the signal response of one read-out channel and
estimate the detector sensitivity when the detector will operate at 50
mK with a complete read-out.  This paper is organized as follows. In
\sref{system_overview} we described the experimental apparatus and in
particular the read-out system. In section \sref{expres} we present
and discuss the experimental results. In \sref{elnoise} and
\sref{eqtemp} we analyzed the electrical system, the noise spectra and
equivalent temperature of the resulting coupled oscillators. Finally
in \sref{strain} we describe the calibration procedure and estimate
the detector strain sensitivity.
   
\section{SYSTEM OVERVIEW}
\label{system_overview}
\subsection{Sphere and mechanical transducers}

MiniGRAIL is a spherical gravitational wave (GW) antenna currently
under development \cite{MINIGRAIL}. The antenna is a massive sphere
in   $CuAl$, has a diameter of $68
\, cm$, a mass of about $1.3 \, ton$ and the GW sensitive spheroidal
quadrupole modes have frequencies around $2980 \, Hz$ at $4.2 \, K$.
The alloy $CuAl 6\% $ has been chosen because of the high quality 
factor ($Q\sim 10^7$ at low temperature), high sound velocity 
($V_S\simeq 4100 \,m/s$) and a sufficient thermal conductivity, 
which allows to cool a 1.3
ton antenna to a  temperature below $ 100 \, mK$ \cite{deWaard04}.  The
ultimate goal is to operate MiniGRAIL at a thermodynamic temperature
of 20 mK, equipped with six transducers coupled to nearly quantum
limited double-stage SQUID amplifiers
\cite{Falferi06,Podt05,Vinante02}. 
 The sphere is suspended from the centre with a gold-plated copper rod
20 mm in diameter. The rod is connected to the last mass of the
mechanical vibration isolation system which consists of seven
mass-spring stages suspended with stainless steel cables from three
absorbers, each consisting of a stack of rubber and aluminium plates.
A detailed description of the detector mechanics and cryogenics can be
found in \cite{MINIGRAIL04}

We used  capacitive transducers to read-out the spheroidal modes. They
consist of  a closed membrane  with a load mass in the
centre. The electrode is made of a thin CuAl plate placed in front of
the resonating mass.
To obtain a small gap between resonator and electrode, we proceeded as
follows. The resonator and the electrode are lapped and
polished to get a smooth flat surface. Further a clean Kapton foil of
a thickness equal to
the desired gap is placed between the electrode and the mass. Finally
a small amount of glue is added between the electrode and the support
springs. A load is applied on top of the electrode in order to make a
compact assembly. After the drying period of the glue the Kapton foil
is removed. This technique was shown to be reliable and
reproducible. Gaps of the order of $20 \mu m$ could be
obtained and voltage bias as large as $500 \, Volt$ could be applied
without discharging.  

In \tref{CMT_table} we  summarise  the features of the three
transducers. Each resonator  has an effective 
resonant mass of about 200 g and is tuned mainly to three different 
spherical modes. 
\begin{table}[htbp]
  \begin{center}
  \small
    \begin{tabular}{c c c c }  
    \multicolumn{4}{c} {\bf Closed membrane transducers}  \vspace{0.1
      cm}\\ \hline
     &  transducer 1 &transducer 2&   transducer 3 \vspace{0.1 cm}\\ \hline
    $mass \, [Kg]$ & $0.205$ & $0.153$ & $0.150$  \\ 
    $C_t \, [nF]$ & $ 1.17\pm 0.01$ & $0.70\pm 0.02$& $1.20\pm 0.01$ \\
    $gap \, [\mu \, m]$ & $ 20\pm 2 $ & $35\pm 4$& $25\pm 2$ \\
    $f_{res}\, [Hz]$ & $ 2863\pm 5 $ & $2850\pm 5$& $2878\pm 5$ \\
    $Q$ at $300 \, K$ & $ 1.0 \cdot 10^4$ & $1.1\cdot 10^4$ & $1.0\cdot 10^4$\\
    $Q$ at $77 \, K$ & $ 3 \cdot 10^4$ & $4.8 \cdot 10^4$ & $ 2.0 \cdot 10^4$\\
      \hline
     \end{tabular}
     \caption{Properties of the three $CuAl 6\%$ closed membrane
       transducers. The transducers resonance frequencies has been
       estimated from the tuning procedure at room temperature as
       described in \cite{GottardiPhD}}.
     \label{CMT_table}
  \end{center} 
\end{table}
\subsection{The calibrator}
The calibrator is a  capacitive resonant transducer. The impedance of the 
calibrator, biased with an electric field $E_{cal}$, for each mode of 
resonance $\omega_{m}$, is given by 
\beq{Zcal}
Z(\omega)=\frac{1}{i \omega C_{cal}}\left(1-\frac{C_{cal}E_0^2}{m_m}
\frac{1}{\omega_0^2-\omega^2+\frac{i \omega \omega_0}{Q_m}}\right)
\eeq
The real part is used to estimate the energy of the mode and can be  
derived as follows
\beq{ReZcal}
\begin{array}{c}
Re(Z(\omega))=\frac{\omega_0 E_0^2}{m_m Q_m}\frac{1}{(\omega_0^2-\omega^2)^2+
\frac{\omega^2 \omega_0^2}{Q_m^2}}=\\\frac{A_{m,0}}{(\omega_0^2-\omega^2)^2+
\frac{\omega^2 \omega_0^2}{Q_m^2}}.
\end{array}
\eeq
on the calibrator 
method described here particularly interesting because it is free from
systematic errors, at least to estimate the temperature of the mode.  

\subsection{The read-out system}

\begin{figure}
 \includegraphics[scale=1]{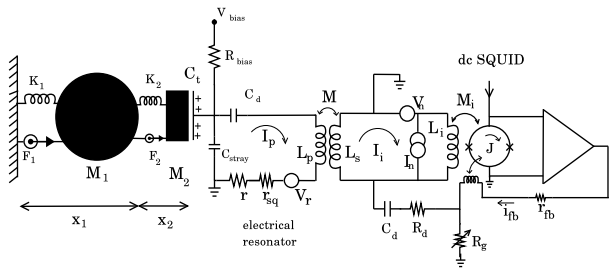}
 \caption{\label{sphere_twomode_cap} Electro-mechanical scheme of a 
spherical antenna with mechanical resonator and capacitive transducer coupled 
to a SQUID through a superconducting matching transformer.}
\end{figure}

Two transducers, named {\it transducer 1} and {\it transducer 2} were
coupled respectively to the two-stage SQUID and to a single stage
commercial Quantum Design SQUID. The third resonator, {\it transducer
3}, was coupled to a room temperature FET amplifier and was used for
diagnostic and calibration purposes. Here we describe the performance
of {\it transducer 1} coupled to a two-stage SQUID system based on a
DROS \cite{Podt99},\cite{Duuren} as an amplifier and a Quantum Design
dc SQUID as sensor SQUID.
\cite{GottardiPhD}. The complete read-out circuit of {\it transducer 1} is
shown in \fref{sphere_twomode_cap}. 
 
The impedance matching between the transducer capacitance and the
SQUID input coil is achieved by using a high-Q superconducting
transformer. The electrical resonance of the transformer is not tuned to
the mechanical modes, so the impedance matching is not optimal.
The transformer coils were made of Nb wires and enclosed into
a double superconducting shield. The bias circuit and the decoupling
capacitor have been  housed into a separate compartment of the
superconducting shielding box.
The measured transformer primary and secondary coil inductances are
respectively $L_p=0.3595 \pm 0.005\, H$ and $L_s=2.1 \pm 0.2  \,\mu H$. The mutual
inductance between the coils was $M=(2.5 \pm 0.5)\cdot 10^{-5} \,H$
and the coupling factor $\alpha_{p,3}= 0.43 $.

We use a decoupling Teflon commercial capacitor $C_d= 220 \, nF$ and
SMD bias resistors for a total resistance of $R_{bias}= 13 \, G\Omega$
at $4.2 \, K$.  The final intrinsic electrical quality factor, after
connecting the decoupling capacitor $C_d$ and the bias resistor
$R_{bias}$, is equal to $Q=(1.8 \pm 0.1) \cdot 10^5$ at $4.2 K$.
 
The transducer has been assembled with a gap of about
$20\pm 2 \, \mu m$ and a capacitance $C=1.17\pm 0.01 \, nF$ measured at room
temperature. This has been done in order to keep the electrical mode 
separated from the mechanical ones. The electrical mode resonance
frequency measured on the antenna at $5 K$ was $8120  \, Hz$.

Two-stage SQUID systems are developed in order to reduce the noise of
dc-SQUID amplifiers,which is normally limited by the room temperature
electronics \cite{Podt99}
\cite{Mezzena01} \cite{Harry01} \cite{Carelli98}. When
used in the transducer chain for resonant gravitational waves detectors,
they can improve the detection sensitivity by orders of magnitude
\cite{Vinante02}. The system described here differs from other two-stage SQUIDs
used in GW experiments since it  uses  a DROS as an amplifier SQUID
\cite{Duuren}. A DROS has a large flux-to-voltage transfer function
which allows direct read-out of the signal. Direct read-out simplifies
multichannel read-out as needed in spherical gravitational wave
detectors.

The two-stage SQUID system we developed is based on a configuration
reported in \cite{Podt99,Gottardi04}.  A standard dc-SQUID chip
manufactured by Quantum Design (QD) \footnote{Quantum Design, 11578
Sorrento Valley Road, San Diego} was chosen as sensor SQUID because of
a larger input inductance with respect to the dc-SQUID described in
\cite{Podt99}.

The dc-SQUID is biased at a constant voltage by means of a resistor $R_{bias}
=1.5 \Omega$. The current through the sensor SQUID is modulated by an applied
signal flux $\Phi_{sig}$ and is fed through the input coil of the DROS. 
The total additive flux noise at $4.2\, K$ is $\sqrt{S_\Phi}=1.60 \pm 0.02
\,\mu \Phi_0 /\sqrt{Hz}$ with input coil open and $\sqrt{S_\Phi}=1.10 \pm 0.02
\,\mu \Phi_0 /\sqrt{Hz}$ with input coil superconductily shorted. This
corresponds respectively to an intrinsic uncoupled energy resolution
of  $\epsilon= S_\Phi/2L_{sq}=650 \pm 15 \, \hbar$ and $\epsilon =320
\pm 15 \, \hbar$. This is in agreement with the expected values
calculated from the SQUID parameters using the standard model \cite{TESCLA77}.
In order to avoid instability in the SQUID-resonator system we
implemented a capacitive cold damping network in the feedback
line. Damping network has been first investigated by Stevenson
\cite{Stevenson96}, using a phase-shifted inductive feedback, and by
Vinante \cite{Vinante02}, who made use of a capacitive network. The
two-stage SQUID coupled to a high quality factor electrical resonator
showed the same performances \cite{GottardiPhD}. We estimate at $4.2\, K$ a
SQUID noise temperature  $T_N=100\pm 30\mu K$ and a noise number
$N=730\pm 100  $. The additive coupled energy resolution was $650
\hbar$ and $320 \hbar$ respectively at $4.2$ and $2.1 \, K$.

\section{EXPERIMENTAL RESULTS}
\label{expres}
\subsection{Electrical system and noise spectrum}
\label{elnoise} 
 The additive noise level of the two-stage SQUID coupled to the
transducer mounted on the sphere, was comparable with the one measured
with the SQUID with open input.
 When operating without the cold damping network, the minimum wideband
 flux noise observed with the SQUID was of $\sim 2.7 \, \mu
 \Phi_0/\sqrt{Hz}$. When the cold damping was  active, we  measured a
 additive wideband flux noise of  $1.67 \pm 0.03 \, \mu
 \Phi_0/\sqrt{Hz}$. It corresponds to an additive coupled energy
 resolution of $730\pm 100 \, \hbar$.  
 \begin{figure}[htb]
 \includegraphics[scale=0.5]{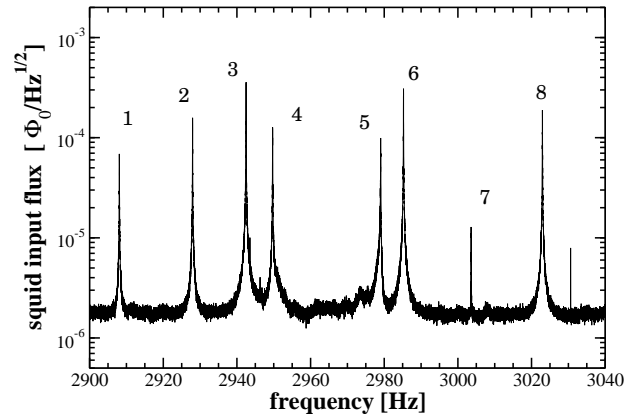}
\caption{Flux spectral density measured at the SQUID output with
transducer bias at $200 \, Volt$. All the expected 8 modes of the
system are visible in the spectrum.}
\label{fluxsquidmodes}
\end{figure} As shown in \fref{fluxsquidmodes}, all the expected 8
modes of the system are visible in the spectrum. To establish which
modes were strongly coupled to the transducer with the two-stage SQUID
system, we proceeded as described in the next section.


\subsection{Equivalent temperature of the coupled oscillators}
\label{eqtemp}
The best and most direct way to estimate the noise of the system is to
measure the input impedance of the transducer coupled to the read-out
SQUID amplifier as described in \cite{Baggio05,Gottardi06}. However,
in this experimental test we did not implement in the matching
transformer the necessary calibration coil.  To estimate the
temperature of the modes we can proceed as follows. The power spectral
density at the output of the calibrator, when the back-action
contribution of the room temperature amplifier is negligible, is
\beq{SVcal} S_{V,CAL}=4k_B T_{eq} Re(Z(\omega)), \eeq where
$Re(Z(\omega))$ was derived in \eref{Zcal}, $T_{eq}$ is the equivalent
temperature of the modes and we are considering monolateral spectra.
At low temperature this value is too small to be measured with the
room temperature FET amplifier. However, by exciting each modes at
resonance with an auxiliary piezo-electric transducer (PZT), we can
increase the signal at a level that can be measured
by the FET amplifier. Then we read both the responses of the FET amplifier and 
the two-stage SQUID coupled to the resonator biased at a voltage $V_{b,sq}$. 
We assume that the voltage $V_{CAL}(\omega)$ at the output of the FET is 
proportional to the voltage $V_{SQ}(\omega)$ at the output of the SQUID, i.e. 
$V_{CAL}=A_{cal,sq}V_{SQ}(\omega)$. This is true at resonance. If we now measure
 the power spectrum $S_{V,SQ}$  at the SQUID output, when the modes are not 
excited,   we can  evaluate the equivalent temperature of the modes using  
\eref{SVcal}, where we substitute 
 \beq{SvASsq}
S_{V,CAL}=A_{cal,sq}^2 S_{V,SQ}.
 \eeq  
 
 \noindent Here  we assumed that the  system response is  linear in the
whole  range.  Linearity has  been  checked  for different  excitation
voltage. We found a linear behaviour within $10\% $.

To estimate  the equivalent temperature  of the modes we  measured the
variance $\sigma^2$  of the stochastic  process with a  spectral noise
$S_{V,SQ}$ at the output  of the two-stage SQUID.  From \eref{ReZcal},
\eref{SVcal}  and  \eref{SvASsq},  defining  $A_{m,0}=  \frac{\omega_0
E_0^2}{m_{m}Q_m}$, the variance can be written as follows
\beq{variance_SQUID}
\sigma_m^2=\frac{4k_BT_{eq}}{2\pi}\frac{A_{m,0}}{A_{cal,sq}^2}
\int_{(\omega_m-\Delta)}^{(\omega_m+\Delta)}\frac{d\omega}{(\omega_0^2-
\omega^2)^2+ \frac{\omega^2 \omega_0^2}{Q_m^2}}, \eeq
 
\noi  from where  we obtain  the relation  which links  the equivalent
temperature of the mode to the variance of the stochastic process with
spectral noise given by $S_{V,SQ}$. We have
\beq{Teq_sigma}     
T_{eq}=\frac{\omega_m^3     A_{cal,sq}^2}{2    k_B
A_{m,0}Q_m}\sigma_m^2=\alpha_m \sigma_m^2, 
\eeq

\noi where $A_{cal,sq}^2$ comes from the calibration as described in
this section, and $A_{m,0}$ is estimated from the Lorentzian curve
fitting of each resonance of the real part of the calibrator
impedance.

To estimate $\sigma_m$ a lock-in amplifier is used with the reference
set at the resonance frequency of the mode. The lock-in amplifier
output magnitude $r$ and angular phase $\phi$ are then sampled at
regular time intervals.  The amplitude decay time constant of the
lock-in amplifier is chosen equal to the sampling time
$\tau_s=\tau_{lk}$. To observe the free evolution of the mode $m$ the
lock-in amplifier time constant is chosen smaller than the time
constant of the mode, $\tau_{lk}<\tau_m$, but large enough that the
lock-in amplifier works as a bandpass filter and makes the
contribution of the broadband noise of the SQUID and the tails of the
neighboring modes negligible.  The mean square amplitude $\langle r^2
\rangle$ of the lock-in amplifier input signal magnitude, equal to the
variance of the total narrow-band noise $V_{nb}^2$, is given by
\cite{BONIFAZI78} \beq{rsq_sigma} \langle r^2
\rangle=V_{nb}^2=\left(1+\frac{\tau_{lk}}{\tau_m}\right)
\left(\sigma_0^2-\frac{S_{wb}}{2\tau_{lk}}\right), \eeq where
$\sigma_0^2$ is the variance of the power spectral density output, and
$S_{wb}$ is the power spectral density of the SQUID wideband
noise. Generally the factor $\tau_m/(\tau_m+\tau_{lk})\sim 1$ in our
case.

 The stochastic process $r^2$ is the sum of two independent Gaussian
processes, the in-phase and quadrature lock-in amplifier output. If
the signal is absent or in general if its average contribution is
negligible with respect to the noise, the variable $r^2$ will have the
exponential distribution
$F(r^2)=\frac{1}{2\sigma_0}e^{-\frac{r^2}{2\sigma_0}}$.

The estimate of $\sigma_0^2$, is then performed by sampling the magnitude 
$r$ at regular time intervals, with sampling time $\tau_s << \tau_m$. A subset 
of data is created by  extracting a data point every resonator time constant 
$\tau_m$ in order to get uncorrelated samples. 
After a large number of samples is collected, a histogram $N(r^2)$ is built, 
where N is the number of samples in a given bin around $r^2$. The histogram is
  fit with the exponential distribution described above and the mean square 
amplitude $\langle r^2 \rangle$ is then extracted as fitting parameter. 

In the absence of excess  or amplifier back-action noise, the quantity
$\langle  r^2  \rangle$ is  proportional  to  the thermal  vibrational
energy in the antenna mode. The constant of proportionality $\alpha_m$
was  used to  rescale the  recorded values  of $\langle  r^2\rangle$ to
antenna energy.  For  the two most coupled modes  at frequencies $2943
\, Hz$ and $2985 \, Hz$  we found the calibration factor $\alpha_m$ to
be    $\alpha_{2943}=(7.0\pm    1.5)\cdot    10^8\,    [K/V^2]$    and
$\alpha_{2985}=(1.1\pm 0.2)\cdot 10^9\, [K/V^2]$ respectively.

\begin{figure*}
\includegraphics[scale=0.8]{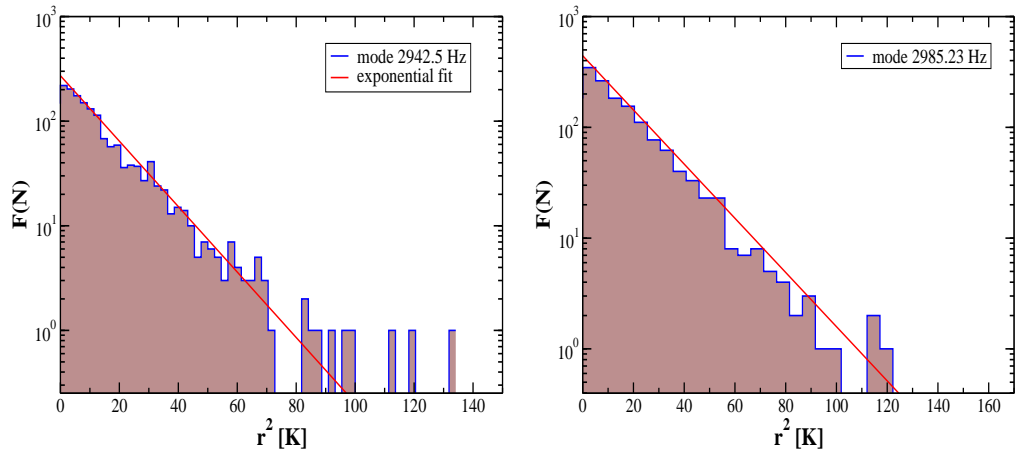}
\caption{\label{distribution}  Exponential distributions  of  the mean
square amplitude $\langle  r^2 \rangle$ for the modes  at $2943 \, Hz$
and $2985 \,  Hz$. The variance $\sigma_m^2$ obtained  from the fit of
the exponential  distribution gives  an equivalent temperature  of the
two modes of $7.0 \pm 2 \, K$ and $9 \pm 2 \, K$ respectively}
\end{figure*}

Graphs (a) and (b) in \fref{distribution} show the energy distribution
estimated for the  modes at frequencies $2943 \, Hz$  and $2985 \, Hz$
during three hours of acquisition. The equivalent temperature for both
modes is obtained by fitting the exponential distribution.
 
 The slope  of the distribution  corresponds to a temperature  of $7.0
\pm 2 \,  K$ for the mode at $2943  \, Hz$ and $9 \pm  2 \, K$ for
the  mode  at  $2985  \,  Hz$,  the  error  arising  mainly  from  the
calibration uncertainty.  The equivalent temperatures  are consistent,
within two sigma, with the  thermodynamic temperature of the sphere.
  No significant difference is
observed in the  equivalent temperature of the mode  between night and
day  acquisitions.   
  
\subsection{Force calibration and strain sensitivity}
\label{strain}
When the mode $m$ is excited at resonance $\omega_m$ with an energy given
 by $\frac{1}{2} k_B T_m$, the power spectral density measured at the 
SQUID output is

\beq{Svsqm}
S_{v,sq}=\frac{4 k_B T_m Re(Z(\omega_m))}{A_{cal,sq}^2} 
\quad \left[\frac{V^2}{Hz}\right],
\eeq
 
\noindent where $Re(Z(\omega))$ is the real part of the calibrator impedance. 

The force power spectral density of a mode excited at 
a temperature $T_m$, when back action is negligible like in our case, is 
given by

\beq{SFFm}
S_{FF,m}=\frac{4 k_B T_m m_{eff,m} \omega_m}{Q_m} \quad 
\left[\frac{N^2}{Hz}\right],
\eeq 

\noindent where $m_{eff,m}$ is the  effective mass of the mode $m$, which
can be  estimated from the tuning  curves of the  sensor transducer if
the bias voltage and the gap are known. \footnote{In a capacitive
transducer the bias electric field introduces a negative spring
constant which causes the resonant frequency of the transducer
resonator to shift down. In the simplest case, this shifted frequency
is related to the transducers electro-mechanical coupling coefficient
$\beta$ by $\beta=- \frac{\omega_t^2-\omega_0^2}{\omega_t^2}$
\cite{Paik76}. $\beta$ is defined as the ratio between the mechanical
and the electrical energy, and describes the conversion efficiency of
mechanical motion into electrical signal. It can be written as follows
\cite{Richard84} $ \beta=\frac{1}{d_0^2}\frac{C_t V^2}{\omega^2
m_{eff}}$ \noi where the effective mass $m_{eff}$ includes possible
geometrical factors \cite{RapOgawa} as well as electrical parameters
from the coupled circuit. See \cite{GottardiPhD} for more details on
the tuning procedure.}  
From the ratio of \eref{Svsqm} and \eref{SFFm} we get

\beq{SvoverSF}
\frac{S_{v,sq}}{S_{FF,m}}=\frac{Q_m  Re(Z(\omega))}{m_{eff,m} \omega_m
A_{cal,sq}^2} \quad \left[\frac{V^2}{N^2}\right].  
\eeq

From the calibrator impedance measurement, we have

\beq{ReZm}
Re(Z(\omega))=\frac{A_{m,c}}{(\omega_{m,c}^2-\omega^2)^2+
\frac{\omega^2\omega_{m,c}^2}{Q_{m,c}^2}}
\eeq

\noi  where  $\omega_{m,c}$ and  $Q_{m,c}$  were  measured during  the
calibration  process and  were different  from the  resonance measured
during  the acquisition  of the  mode spectra  because, in  the later
case, the calibrator  was not charged. $A_{m,c}$ is  obtained from the
Lorentzian  fit  of each  mode  of the  real  part  of the  calibrator
impedance.

At each mode resonance \eref{SvoverSF} becomes

\beq{SoverSF_res}
\left(\frac{S_{v,sq}}{S_{FF,m}}\right)_{\omega=\omega_m}=
\frac{Q_m A_{m,c} }{m_{eff,m} \omega_m A_{cal,sq}^2}
\frac{Q_{m,c}^2}{\omega_m^4} \quad \left[\frac{V^2}{N^2}\right].
\eeq 

We now turn to analyze the detector transfer functions.
The relations derived so far are valid at resonance. As a first
approximation, one can consider the transfer function ${\cal
G}_{SQ,F}$ as a product of poles and zeros where the poles are derived
from the polynomial fit of the SQUID noise spectrum and the zeros are
 chosen to fit the measured amplitude at resonance  given by
\eref{SoverSF_res}.

The transfer functions for an applied calibration  signal becomes
\beq{Gsqf}
{\cal
G}_{SQ,F}(\omega)=H_{m,cal}(\omega)\frac{\Pi_{k=1}^{N_r} 
(j\omega-r_{k,m})(j\omega-r^\ast_{k,m})}{\Pi_{k=1}^{N_p}
(j\omega-p_k)(j\omega-p^\ast_k)}.
\eeq   
In the equation above  $N_p>N_r$ and $H_{m,cal}(\omega)$  is a force 
calibration constant which has been experimentally determined from the
calibration measurement at resonance.

 All the terms included in \eref{SoverSF_res} and \eref{Gsqf} are derived
experimentally from the calibration, from the tuning curves and from
direct measurement of the modes quality factor. The transfer function
is experimentally measured at resonance and only approximated out of resonance. 

 \noindent We remark again that quality factors and
resonance frequencies are different when measured during 
calibration $(\omega_{m,c},Q_{m,c})$ and during  acquisition of the
noise spectra $(\omega_{m,s},Q_{m,s})$, due to the bias voltage in
the calibrator. This complication arises from the fact that the
calibrator was also coupled to the quadrupolar modes of the
sphere. This effect is included in the estimate transfer function of
\eref{Gsqf}.

A small parenthesis needs to be opened here. While for a bar detector
 it is relatively straight forward to relate the strain produced by a
calibrator located on one of the bar faces with the strain from a
gravitational wave signal, the same cannot be said for a spherical
detector. A calibration {\it hammer stroke} excites a linear
combination of the five spheroidal modes depending on the position on 
the sphere surface. When only one calibrator is used, like for the
Minigrail test run described in this paper, one only calibrates the
detector for a particular set of forces $\mathbf{F_m}$ applied to each
spheroidal mode. Such a  combination of forces might not always
represent a GW excitation. In order to fully calibrate the detector
one needs to generate calibration forces from a set of 5 or more
calibrators located at different positions on the sphere surfaces.
 A detailed procedure to calibrate a spherical
detector is described in  \cite{Gottardi06}.
 
In the experiment described here, the calibration signal is generated
by a piezo located at position $(\theta = 18^{\circ} ,\phi
=135^{\circ}) $. By using Eq. 45 in \cite{Gottardi06}, derived
previously in \cite{Stevenson98,Merkowitz98,Lobo00}, one finds that a
piezo in such a location excites a combination of the 5 spheroidal
modes given by the vector $(0, 0.13, -0.49,0.5, -1)$, normalised to
the maximum value of its elements. One can see that such a combination
is equal, within $20 \%$ tolerance, to the one generated by a
circularly polarised gravitational wave coming from direction
$(\theta=20^{\circ} ,\phi=135^{\circ} )$. Following \cite{Gottardi06}
one can see that such a wave direction is almost optimal for a
transducer configuration described in this paper.
 
\begin{figure}[htb]
\includegraphics{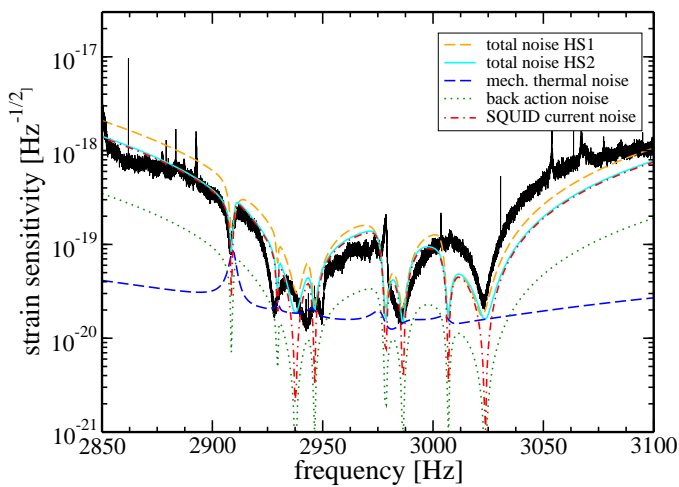}
\caption{MiniGRAIL strain sensitivity at $5 \, K$ with three
transducer placed on the sphere, but single transducer read-out.The
sensitivity has been estimated for a particular combination of the 5
spheroidal modes as derived by exiting the mode using a piezo
transducer (PZT) located at $(\theta=18^\circ ,\phi= 135^\circ)$.  The
dashed gray curve shows the strain sensitivity calculated for a
simulated hammer stroke excitation from the PZT location using the
electro-mechanical model described in \cite{Gottardi06}. For the
simulation we used the detector parameters discussed here in the text.
A better matching between the experimental data and the simulation is
obtained when the simulated hammer stroke is given at the position
$(\theta=27^{\circ} ,\phi=135^{\circ})$ (continuous gray curve).  The
other curves shows the contribution of the thermal noise (dashed dark
line) and the back-action noise (dotted line) to the strain
sensitivity}
\label{strain5K200V}
\end{figure}

MiniGRAIL strain sensitivity with three transducers coupled to the
spheroidal modes, but only one used for the read-out, is shown in
\fref{strain5K200V}. The read-out transducer was biased with an
constant electric field of $E=10^7 Volt/m$ and the sphere
thermodynamic temperature was $5.2 \, K$.
As discussed above, the experimental strain curve gives and estimation
of the detector sensitivity only for a particular combination of
spheroidal modes corresponding to a gravitational wave coming from the
$(\theta=20^{\circ} ,\phi=135^{\circ})$ direction.

We obtained a peak strain sensitivity of $(1.5\pm 0.6)\cdot 10^{-20}
Hz^{-1/2}$ at $2942.9 \, Hz$ and a strain sensitivity of about $5\cdot
10^{-20} Hz^{-1/2}$ over a bandwidth of $30 \, Hz$. This corresponds to
a strain amplitude of $h \simeq 2.5\cdot 10^{-18}$ at $3 \, kHz$ for a
burst signal of $1\, ms$ \cite{COCCIA96}. For a sphere of $68 \, cm$
in diameter like the one of MiniGRAIL, it is equivalent to a
displacement sensitivity, at $3 \, kHz$, of $1.6 \cdot 10^{-19} \, m$.
When optimal filtering is applied to the output signal, the detector
is sensitive to burst signals with an impulse energy of about $T_N\sim
50 \, mK$' as can be derived calculating the noise temperature using
the experimental data. This sensitivity
would be enough to detect supernova explosions in our galaxy.  The
calculated sensitivity curves  in \fref{strain5K200V}, are obtained
using the model described in \cite{Gottardi06}.
 The dashed gray curve
shows the strain sensitivity calculated for a simulated  hammer-stroke
excitation applied to the same point on the sphere where the
PZT is placed, i.e. 
$(\theta=18^{\circ} ,\phi=135^{\circ})$. The simulated signal describes
reasonably well the strain sensitivity 

The best fitting strain sensitivity has been
obtained for a simulated hammer stroke applied at the sphere surface  point
$(\theta=27^{\circ} ,\phi=135^{\circ})$. The result is shown with the
continuous gray line. In this case the third mode
is more excited. The agreement with the experimental data is
rather impressive considering the amount of fitting parameters involved
in the simulation. The difference of about
$7^\circ$ in the angle $\theta$ of the experimental and simulated excitation
position  could be explained considering the fact that a spherical detector with
only three transducers in the position 1,2 and 5 like the one considered
here is far from being symmetric. In \cite{Merkowitz97} the authors
had to perform  a rotation of spheroidal mode reference frame as well
to be able to explain their experimental results. In their case the
transducers were not as massive as here and, above all, they used six
transducers  positioned in the symmetric TI configuration.
The resulting  mixing of the spheroidal modes could explain
the discrepancy between the measured and the simulated sensitivity curves.   
In order to address more accurately this issue one should place on the
sphere  at least 5 calibrators to measure the detector response to
each of the 5 spheroidal modes. When six transducer are fully operating
with comparable sensitivity, a single calibrator is enough to fully
calibrate the detector as shown in \cite{Merkowitz97,Gottardi06}.

The contribution to the strain sensitivity of the well known noise
sources are plotted in \fref{strain5K200V} as well. At resonance, the
sensitivity is limited by mechanical thermal noise of the transducer
mass. Out of resonance the sensitivity is limited by the SQUID
additive current noise. The back action noise of the SQUID is about an
order of magnitude smaller. The electrical thermal noise of the
superconducting transformer, not shown in the graph, is negligible
because the electrical mode is well decoupled from the mechanical ones. 

From the measurement of the variance of the most coupled modes and the
simulated data we can conclude that, within the experimental accuracy,
the Minigrail peak sensitivity is currently limited by the thermal
noise generated by the transducer mass. Some of the modes, however,
show excess noise whose origin is difficult to address. A better
estimate of the transfer function of each spheroidal mode is necessary
in order to fully characterise the detector. The measurements
presented here have to be considered as a first test bed for the
following engineering and science runs of Minigrail.  The sensitivity
is expected to improve of at least one order of magnitude when the
detector will operate at 20 mK.

\begin{figure}[htb]
\includegraphics{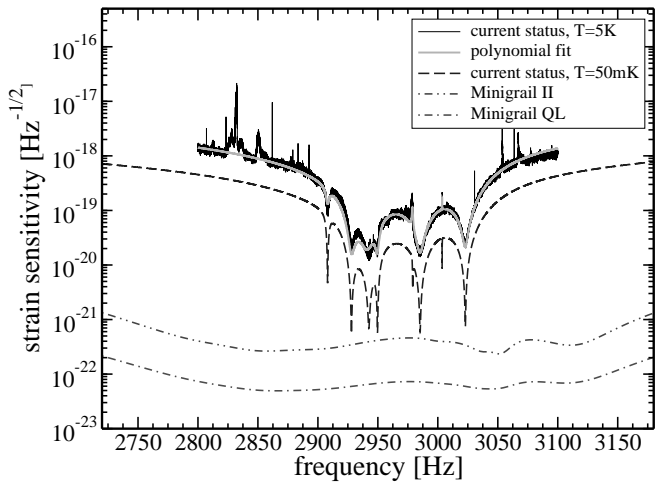}
\caption{The measured strain sensitivity of MiniGRAIL is shown
together with predicted sensitivity for future detector
configurations. The continuous gray line is a polynomial fit of the
measured strain sensitivity. The dashed line shows the sensitivity for
the detector operating at $T=50 \, mK$ with the same three transducers
configuration presented in this paper.  The dot-dot-dashed line
(Minigrail II) shows the sensitivity achievable with available
technology , namely $T/Q\sim 2.5\times 10^{-8}\,K$ and SQUID energy
resolution $E=70 \hbar$. The dot-dashed curve gives the sensitivity
for a quantum limited detector (Minigrail QL) with $T/Q \sim 1\times
10^{-9}\,K $ .}
\label{strainfuture}
\end{figure}

 In \fref{strainfuture} the measured strain sensitivity is shown
 together with predicted sensitivity for possible future detector
 configurations. A polynomial fit of the strain sensitivity is shown
 as well. The poles and zeros obtained from the fit can be used to
 build the matched filters for GW detection \cite{Gottardi06}.
 The expected strain sensitivity of Minigrail is shown for the detector
 operating at  $T=50 \, mK$ with the same three transducers configuration
 presented here.  The figure shows the expected sensitivity for a
 Minigrail II. In this configuration Minigrail operates with 6
 capacitive transducers placed in the TI configuration where the
 electrical modes are coupled to the mechanical ones. We consider 
 $T/Q=2.5\times10^{-8} K$ and a SQUID coupled energy resolution
 $E_{coupled}=70 \hbar$. Those values are achievable within the current
 technology. See \cite{Gottardi06} for a recent review. The sensitivity
 of Minigrail operating at the quantum limit is also shown.
 Minigrail can reach a peak sensitivity of about $6\cdot 10^{-23}
Hz^{-1/2}$ and a bandwidth larger than 400 Hz at a sensitivity of
$1\cdot 10^{-22} Hz^{-1/2}$.

\section{CONCLUSIONS}
\label{conclusions}

We have operated at 5 K a spherical resonant detector equipped with
a capacitive resonant transducers coupled to a two-stage SQUID
amplifier. Our two-stage SQUID amplifier is one of the most sensitive
amplifier employed so far on a GWs resonant detector. We measured an
additive coupled energy resolution of  $700\pm 100\, \hbar$ at $5 K$   
We reach a peak  strain sensitivity of
$1.5\cdot 10^{-20}Hz^{-1/2}$ at 2942.9 Hz. A strain sensitivity of
better than $5\cdot 10^{-20}Hz^{-1/2}$ has been obtained over a
bandwidth of 30 Hz.  We expect an improvement of more than one order of
magnitude when the detector will operate at 50mK. This result should
be considered as the first step towards the realization of an
ultracryogenic spherical gravitational wave detector.

\begin{acknowledgments}
The authors would like to acknowledge precious discussions with
Jean-Pierre Zendri, Paolo Falferi, Andrea Vinante and Alberto Lobo. We
are grateful to Hibbe van der Mark for his technical help. This work
has been partially financially supported by Integrated Large
Infrastructures for Astroparticle Science (ILIAS) of the Sixth
Framework Programme of the European Community.
 
\dots.
\end{acknowledgments}

\bibliography{GOTTmGrail}

\end{document}